\documentclass[12pt]{elsarticle}
\usepackage{graphicx}
\usepackage{caption}
\usepackage{subcaption}
\biboptions{authoryear}
\begin{document}

\title{Operational risk of a wind farm energy production by Extreme Value Theory and Copulas}
\author{Guglielmo D'Amico}  
\address{Dipartimento di Farmacia, 
Universit\`a `G. D'Annunzio' di Chieti-Pescara,  66013 Chieti, Italy}
\author{Filippo Petroni}
\address{Dipartimento di Scienze Economiche ed Aziendali,
Universit\`a degli studi di Cagliari, 09123 Cagliari, Italy}
\author{Flavio Prattico}
\address{Dipartimento di Ingegneria Industriale e dell'Informazione e di Economia, Universit\`a degli studi dell'Aquila, 67100 L'Aquila, Italy}

\begin{abstract}
In this paper we use risk management techniques to evaluate the potential effects of those operational risks that affect the energy production of a wind farm. 
We concentrate our attention on three major risk factors: wind speed uncertainty, wind turbine reliability and interactions of wind turbines due mainly to their placement.   

As a first contribution, we show that the Weibull distribution, commonly used to fit recorded wind speed data, underestimates rare events. Therefore, in order to achieve a better estimation of the tail of the wind speed distribution, we advance a Generalized Pareto distribution. The wind turbines reliability is considered by modeling the failures events as a compound Poisson process.
Finally, the use of Copula able us to consider the correlation between wind turbines that compose the wind farm. Once this procedure is set up, we show a sensitivity analysis and we also compare the results from the proposed procedure with those obtained by ignoring the aforementioned risk factors.


\end{abstract}


\maketitle

\section{Introduction}
Wind power now accounts for a high proportion of generation capacity in many regions. For example, wind accounted for 17 percent of Germany's 167.8 GW of installed generation capacity in 2011 (\cite{hau2013wind}).
As wind power's share of electricity generation has increased, so have the financial consequences of risks associated with its inherently high variability. 
Wind speed variability has many financial consequences. Low wind speeds reduce generation and revenues for wind power generators and may adversely affect the ability to meet debt payments, creating credit risks for investors. 
Conversely, excessively high wind speeds may temporarily halt generation or delay wind farm construction. 
When wind has priority access to the grid, thermal power plants have to balance generation regardless of whether wind is above or below forecasted levels. Wind speed variability may also compound price risk for other market players through its influence on wholesale electricity market clearing prices in competitive day-ahead and intraday markets. 
The uncertainty in wind power production needs to be hedged trough risk management techniques. 

In this work we will focus our attention on the source of operational risk which are present in a wind farm, namely: wind speed uncertainty, wind turbine reliability and interactions of wind turbines due mainly to their placement. 

Despite the Weibull distribution (WD) is often used by practitioners and researcher alike (see \citet{weis03,akda09,chan11}) it does not fit well the right tail of the wind speed distribution underestimating strong wind probabilities. The WD models accurately the body of the wind speed distribution, but the same statement can not be made for the tail of the distribution. By applying extreme value theory we will show that it is possible to better estimate the number of strong wind events if a Generalized Pareto distribution (GPD) is used to fit the right tail of the wind speed distribution. A similar approach was already used in the field of the wind speed modeling (see \citet{morg11,holm99,van11,zach98}). Here we are interested into highlighting the importance of extreme value theory as a mean for controlling the operational risk arising from the uncertainty of the wind speed as applied to a real case of energy production. 

Another source of operational risk is the wind turbine reliability  and the necessary time to repair. We model the failure events by means of a compound Poisson process.
The compound Poisson model is widely adopted by insurance modelers for measuring aggregate risks (see e.g. \citet{tse09}) and we will show that it can be used also in the management of a wind farm to consider periods of non production of energy due to failure of the wind turbines and to the time for repairing.

The third source of operational risk we consider is the correlation between wind turbines energy production. Indeed, since in a wind farm many turbines act together, it would be better to consider their multivariate distribution of energy production instead of considering the turbines as independent and with an identical production of energy. This risk factor is considered through Copulas that permit the construction of a multivariate model having fixed marginals (univariate) distributions.

In the present work we consider a wind speed database of a specific site in Alaska and we assume to put there a wind farm composed of 10 commercial wind turbines. We propose then a procedure to estimate correctly the energy production of the allocated wind farm by taking into account the three sources of uncertainty. 

The paper is organized as follow: in section \ref{data} we describe 
the wind speed database and the commercial wind turbines considered in the application. In section \ref{model} we present the models at the basis of the proposed procedure. Section \ref{app} shows the application of the procedure to our database, sensitivity analysis and the comparison with the energy estimation of a wind farm without considering the aforementioned risk factors.  At last, in section \ref{res} we give some concluding remarks.

\section{Material}
\label{data}

\subsection{Database}\label{db}
The database of wind speed used in our analysis was collected by the National Data Buoy Center (www.ndbc.noaa.gov). Particularly, we downloaded the data from the inshore station RDDA2 that is situated at 67.577 N 164.065 W in Alaska. The data are available for six years ranging from 2006 to 2012 with a sample period of six minutes. The instrumentation are located at 10 $m$ above the ground and mean and maximum values of wind speed in the database are respectively of 4.5 and 34.8 $m/s$.

This database is used to analyze the production of energy from commercial wind turbines which have the hub at a given altitude. Then, since the altitude from the ground influences the wind speed, we have to transform the 10 $m$ velocities to corresponding data at the required altitude. It is well known in the literature that wind speed has the following dependence from the altitude (see e.g. \citet{windb}):
\begin{equation}\label{height}
v_h = v_{rif} \left( \frac{h}{h_{rif}} \right)^{\alpha} \;\;\;\;\; \alpha= \frac{1}{ln\frac{h}{z_0}}
\end{equation}
where $v_h$ is the wind speed at the height of the wind turbine hub, $v_{rif}$ is the value of the wind speed at the height of the instrument, $h$ and $h_{rif}$ are the height of the wind turbine and of the instrument ($h=50m$ and $h_{rif}=10m$), respectively. The parameter $z_0$ is a factor that takes into account the morphology of the area near the wind turbine. For a region without buildings or trees, this parameter varies from 0.01 and 0.001, instead for the offshore application it is equal to 0.0001. In our analysis we consider a mean value for an onshore application, then we fix $z_0$=0.005. With this transformation we have an increase of the mean and also of the maximum value of the wind speed, which became 5.4 and 42 $m/s$ respectively. In figure \ref{db} we show the main characteristics of the database. In panel $(a)$ we show the probability density function (PDF) of the wind speed. Panel $(b)$ shows a piece of one year of the time series, instead in panel $(c)$ we report the Box-Plot where we can see that the median wind speed is below the mean value and that in the fourth quartile there are all the wind speeds greater than 15 $m/s$.

\begin{figure}
\centering
\includegraphics[height=9cm]{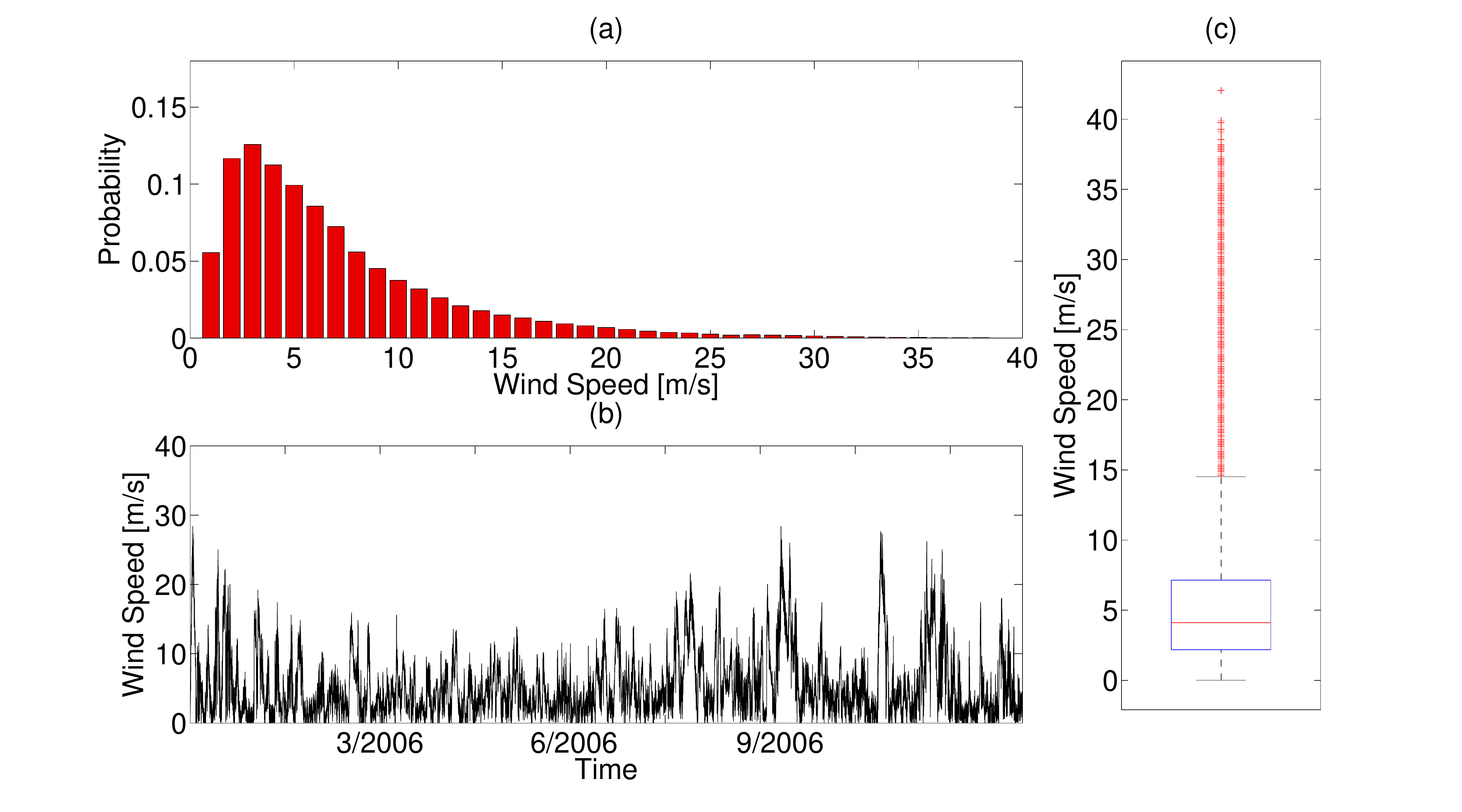}
\caption{Wind speed data of the RDDA2 station. $(a)$ Histogram of the probability density function, $(b)$ one year time series, $(c)$ Box-Plot}\label{db}
\end{figure}

\subsection{Commercial wind turbine}\label{blades}

\begin{figure}
\centering
\includegraphics[height=8cm]{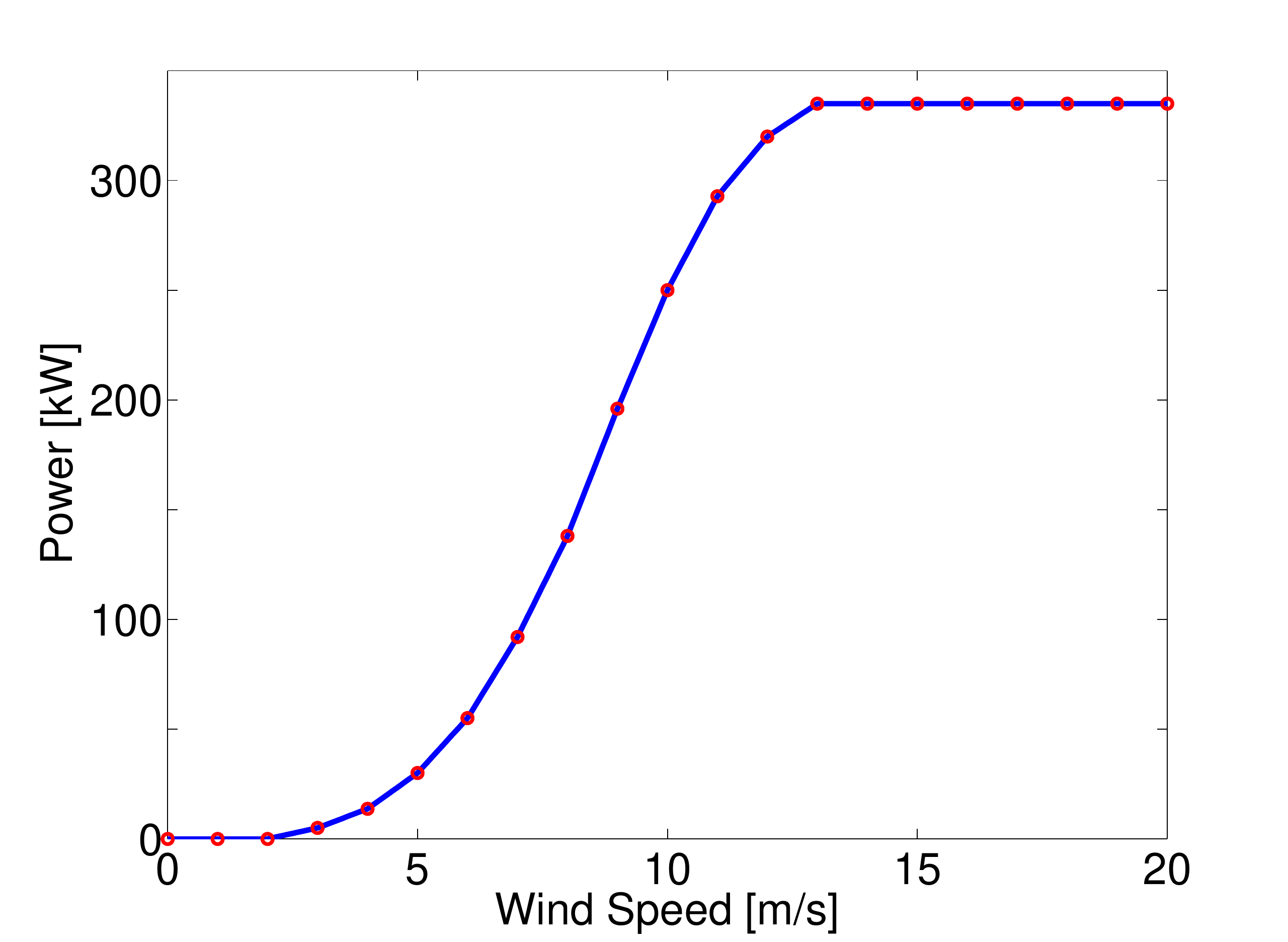}
\caption{Power curve of wind turbine Enercon E33 with a rated power of 330 $kW$.}\label{pc}
\end{figure}

 Wind turbines convert the kinetic energy of wind into electrical power. The quantity of converted energy depends, ceteris paribus, on the installed wind turbines. In this application we chose a commercial wind turbine, the 330 $kW$ Enercon E33. This turbine has an height of the hub from the ground of 50 $m$. The most important property of each wind turbine is its power curve that characterizes the performance of the wind turbine. This curve gives the energy produced by the turbines as a function of wind speed. The power curve of the 330 $kW$ Enercon E33 is represented in figure \ref{pc} and the numerical values are reported on table \ref{tab1}. For the present application given the continuous nature of wind speed, in order to convert each wind speed into energy a linear interpolation between discrete states of the power curve was performed.

\begin{table}
\begin{center}
\begin{footnotesize}
\begin{tabular}{|p{2cm}|c|c|c|c|c|c|c|c|c|c|c|c|c|c|}\hline
Wind speed [m/s] & 0 & 1 & 2 & 3 & 4 & 5 & 6 & 7 & 8 & 9 & 10 & 11 & 12 & 13 $ \div$ 28\\
\hline
Power $\;\;$ [kW] & 0 & 0 & 0 & 5 & 13.7 & 30 & 55 & 92 & 138 & 196 & 250 & 293 & 320 & 335\\
\hline
\end{tabular}
\end{footnotesize}
\caption{Numerical values of the power curve of the wind turbine Enercon E33.}\label{tab1}
\end{center}
\end{table}
The given wind turbine starts to produce energy at a given wind speed (cut-in), below this speed the kinetic energy of the wind is too low to move the blades, viceversa there exist a value of wind speed (cut-off) above which the blades cannot operates and are disposed parallel to the wind speed flow to avoid structural breaks. As it is possible to see in table \ref{tab1} the Enercon E33 has a cut-in wind speed of $3 m/s$ and a cut-off wind speed of 28 $m/s$

\section{Wind speed model}
\label{model}
In this section we describe the mathematical tools used to simulate the energy produced by a wind farm considering the three sources of operational risk. We concentrate the attention on the estimation of the wind speed distribution at a given time interval and not on the adoption of a stochastic process approach to model wind speed as for example in \cite{wind1,wind2,wind3,wind4,wind4b,wind5}.
\subsection{Step 1: fitting the wind speed distribution}
In order to fit the experimental distribution of wind speed many authors suggest the use of a Weibull distribution, see e.g. \citet{weis03,akda09,chan11}. A random variable $X$ has a 2-parameter Weibull distribution if its probability density function obeys relation:  
\[
g(x)= \left( \frac{a}{b} \right) \left( \frac{x}{b} \right) ^{a-1} e^{- \left( \frac{x}{b} \right) ^{a}},
\]
where $a$ is the shape parameter and $b$ is the scale parameter.\\ 
We are going to show that this distribution is not able to take into account the extreme wind speed values and this has important consequences when dealing with energy production. A good way of considering these extreme events is by using appropriate fat tail distributions. In this paper we adopt the generalized Pareto distribution (GPD) as a model for extreme wind speed values. A random variable $X$ has a GPD distribution if its probability density function obeys relation:  
\[
h(x)=\frac{1}{\sigma} \left(1+\frac{\Xi (x-\mu)}{\sigma}\right)^{\left(-\frac{1}{\Xi}-1\right)},
\]
where $\mu$ is the location parameter, $\sigma$ is the scale parameter and $\Xi$ is the shape parameter.\\
Following extreme value theory, we use the WD to model the body of the experimental distribution (ED) and GPD to model the right tail of the ED. Summarizing, we ended up with a threshold density that consider the WD for non-exceeding wind speed and a GPD distribution for wind speed values that fall above the fixed threshold $K$.
Therefore, the considered wind speed distribution is the following:

\begin{equation}
f(x)=\left\{
\begin{array}{cl}
\ g(x) &\mbox{if $x<K$,}\\
  h(x) &\mbox{if $x\geq K$}.
\end{array}
\right.
\end{equation}
The function $f(x)$, as it is defined, it is not a probability function, in fact:
\begin{equation}
\int f(x)dx=z\neq 1
\end{equation}
then $f(x)$ is normalized as follow:
\begin{equation}\label{pdf}
\tilde{f}(x)=\frac{f(x)}{z}  \;\; \forall x.
\end{equation}

\subsection{Step 2: modeling the multivariate wind speed distribution}
The second step of the procedure has the objective of recovering the joint distribution of wind speed for the different installed wind turbines. Let us assume that the wind farm consists of $n$ wind turbines. We denote by $F_{i}(x)$ the cumulative distribution of wind speed at the i-th turbine where $F_{i}(x)=\int_{0}^{x}\tilde{f}_{i}(s)ds$ and $\tilde{f}_{i}(x)$ is given in formula $(\ref{pdf})$. The distributions of wind speed of different turbines installed at a given location are strongly dependent because there are numerous common factors that affect all these distributions such as the underlying wind, geomorphological factors, shear effect and others. Since we have determined at the previous step a suitable marginal distribution (for each single turbine), we desire to maintain it while extending the analysis to model the multivariate distribution. As it is well known, the copula function represent a solution to this problem. 

An n-dimensional copula $C$ is a mapping from $[0,1]^{n}$ to the set $[0,1]$. It is grounded and n-increasing and satisfies the following condition:

The margins $C_{i}$ satisfy $C_{i}(u) = C(1, ..., 1, u, 1, ..., 1) = u$  for all $u\in [0,1]$.   

If we denote by $F_{\mathbf{X}}(\mathbf{x})$ the joint distribution of wind speed at the different turbines, then by Sklar theorem there exists a unique copula $C$ such that

\[
F_{\mathbf{X}}(\mathbf{x})=C(F_{1}(x_{1}), F_{2}(x_{2}),..., F_{n}(x_{n}) ).
\]
and conversely, if we specify a copula function $C$, then $ C(F_{1}(x_{1}), F_{2}(x_{2}),..., F_{n}(x_{n}) )$ is a multivariate distribution function with marginal distribution function $ F_{i}(x_{i})$ for all $i=1,...,n$.

The problem of selecting an appropriate copula function will be discussed later on the applicative section.

\subsection{Step 3: simulation of vectors of wind speed}
\label{gugu}
The third step consists in the simulation of $n$ vectors $(v_{1}(\cdot), v_{2}(\cdot),..., v_{n}(\cdot))$ of correlated wind speed. The vector $v_{i}(\cdot)$ represents the values of wind speed acting on the i-th turbine. This vector has dimension $1\times T$ where $T$ is the horizon time of interest. In this study we fix $T=365\times 24 \times 10$ which represent a six minutes frequency for one year. 
The Monte Carlo simulation starts by generating $n$ time series of a certain length composed of random numbers, ranging from $0$ to $1$, correlated through the given Copula's law. The transformation
of these random numbers into wind speed (following a specific PDF) is made through the inverse of the W-GPD cumulative distribution function previously explained and showed in Figure \ref{cdf}. The application of the proposed procedure has hitherto generated $n$ time series, of length $T$, correlated through the Copula and following the W-GPD.

\subsection{Step 4: modelling wind turbine failures}

As the totality of mechanical device, also wind turbines are subject to failures that can determine the break of the wind energy production. In this step we specify a simple model of failure, namely the compound Poisson model that we will use later for computing the total time of inactivity of the turbines and the consequent loss of energy production (see also \citet{tavn07}).

Let denote by $N$ the number of failures in the block of wind turbines, and by $X_{i}$ the total time of inactivity of the given wind turbine that supported the $i$-th failure. Then the total time of inactivity is given by
\[
S=\sum_{i=1}^{N}X_{i}.
\]

The key and simplifying hypotheses that are commonly done are:
\begin{itemize}
\item The random variables $N$ and $X_{i}$ are assumed to be independent for all $i=1,\ldots,n$;
\item The random variables $(X_{i})_{i=1}^{N}$ are independent and with the same distribution as those of the random variable denoted by $X:\Omega \rightarrow N$.
\end{itemize}

The consequence of these hypotheses is that the random variable $S$ follows a compound distribution, with $N$ being the primary distribution and $X$ the secondary one and then the probability generating function of S is given by:

\[
P_{S}(t)=\mathbf{E}[t^{S}]=P_{N}(P_{X}(t)).
\]

To specify completely the model here below we assume that the distribution of $N$ is a Poisson of parameter $\lambda$ and the distribution $X$ of the repair times is also a Poisson with parameter $\mu$.\\ 

The compound Poisson model is a very popular option adopted by insurance modelers for measuring aggregate risks, see e.g. \citet{tse09} . Many of the advantages of this stochastic model translate naturally to the analysis of wind farm. One of the main advantages is that, if the wind farm is expanded by an increase of the number of installed turbines, then this choice will impact only on the frequency of failures (the primary distribution) but not the time necessary to repair it (the secondary distribution). Differently, innovation in repair management, may affect only the distribution of the time of repair of the turbine and not the frequency distribution. Furthermore, general innovation in technology and quality of materials may affect both the primary and secondary distribution differently, therefore a separate modeling through the primary and secondary distribution allows the comprehension and measurement of the effects of events on the aggregate time of non-production.

\subsection{Step 5: simulation of failures and repair times for the wind farm}

For each one of the $n$ wind turbine we simulate the number of failures and the repair times within the horizon time $T$. Moreover for each failure we simulate the positioning on the time interval $[0,T]$ and then we modify the vectors of correlated wind speed $(v_{1}(\cdot), v_{2}(\cdot),..., v_{n}(\cdot))$ simulated at Step 3 by substituting the velocities with zeros in correspondence of times of repair of failures. This step can be formally described by the following algorithm where comments follows instruction being inserted between parenthesis.\\

\begin{itemize}
\item For $i=1:n$; (for each wind turbines) 
\item Sample $f_{i}$ from $Poisson(\lambda)$; (simulate $ f_{i}$ the number of failures for each turbine)
\item For $j=1:f_{i}$ Sample $r_{i}(j)$ from $Poisson(\mu)$; (simulate $ r_{i}(j)$ the repair time for the j-th failure of the i-th turbines)
\item Set $s_{i}(0)=0$ and Sample $s_{i}(j)$ from $Uniform[s_{i}(j-1)+r_{i}(j-1)+1,T]$; (simulate $s_{i}(j)$ the time of the j-th failure of the i-th turbine. Notice that next failure can occur at time $s_{i}(j+1)$ which lies after the repair of the j-th failure $ s_{i}(j)+r_{i}(j)+1$ and the horizon time $T$) 
\item Set $F_{i}=\sup\{a=1,...,f_{i} : s_{i}(a)\leq T\}$; ($F_{i}$ is the effective number of failures supported by the i-th turbines within time $T$)
\item For all $j=1: F_{i}$
Set $ v_{i}(t)=0$ for all $t\in [s_{i}(h), s_{i}(h)+r_{i}(h)], h=1,2,...,F_{i}$; (this instruction replace the velocities simulated at Step 3 with values equal to zero in correspondence of all times when the turbines are not working due to the failures and the repair times)
\end{itemize}

\subsection{Step 6: computation of the energy production}

As last step we compute the total energy produced by the wind farm using the vector of velocities modified in Step 5. The vectors of velocities (one for each wind turbine) is converted into energy production by using the power curve that characterizes each wind turbine. In this way we obtain the energy produced at each period from each turbine and then summing over all periods and turbines we obtain the energy produced by the wind farm. It should be remarked that to compute the total energy we used dependent vectors of wind speed whose marginal considers appropriately extreme events and a model of failure of the wind turbines. In Section \ref{spie} we will better highlight the main steps of the conversion of the wind speed into energy produced by the wind turbine.

\section{Results}
\label{app}
In this section we will apply the methodology previously explained to the specific wind speed dataset introduced in section \ref{data}. 

\subsection{Parameter estimation of the PDF}
\begin{figure}
\centering
\includegraphics[height=8cm]{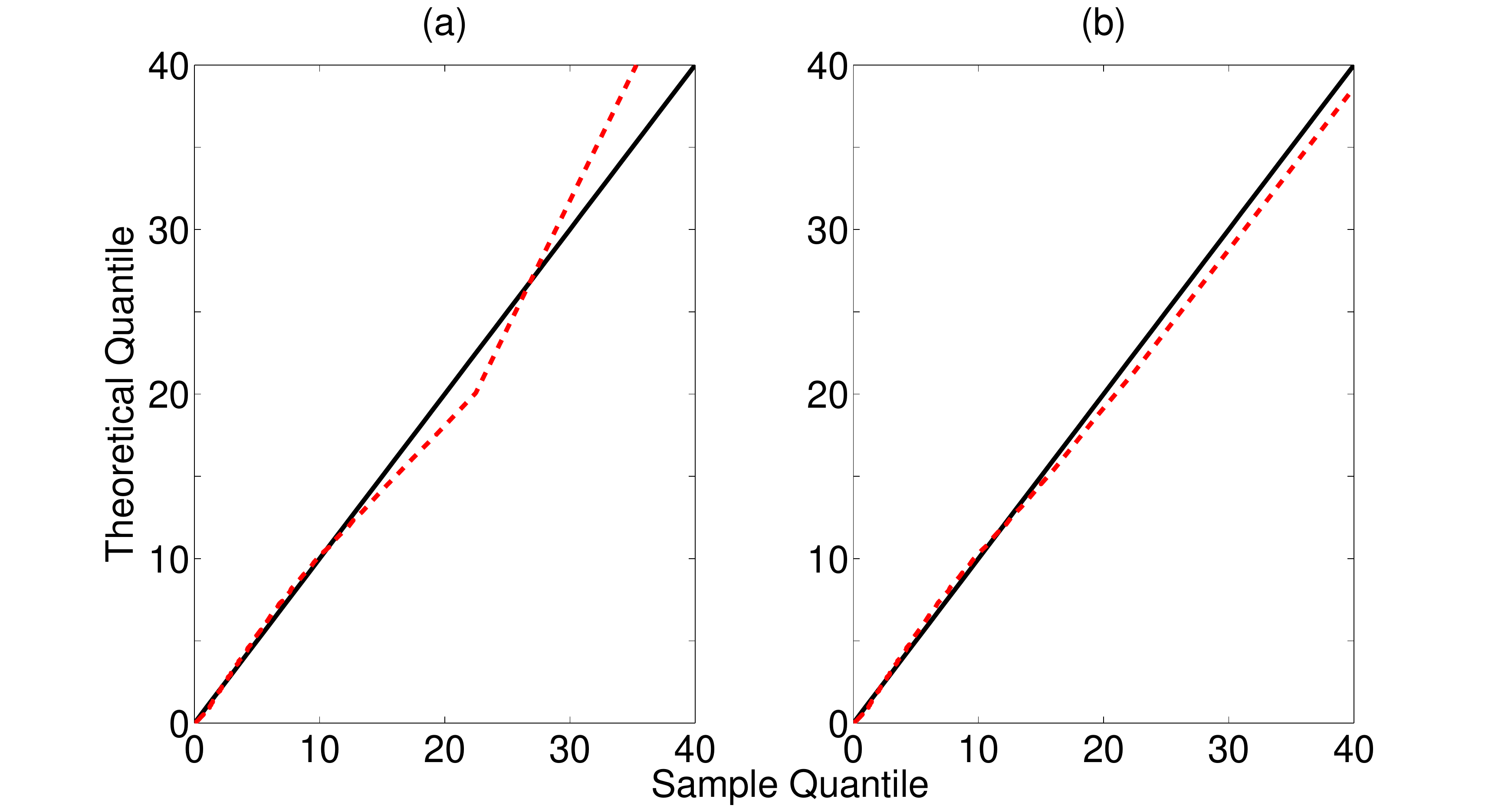}
\caption{Panel (a): Q-Q plot of the WD and the empirical distribution. Panel (b): Q-Q plot of the W-GPD and the empirical distribution.}\label{qq}
\end{figure}

At the basis of the proposed procedure is the possibility to model the PDF of wind speed by two different distributions. While the WD is used to model the first part of the empirical PDF, a GPD is used for the right tail. Then, the first thing to do, in order to apply this methodology, is to choose a wind speed threshold under which we use the WD and over the GPD. 
We then fit using maximum likelihood estimation a WD to our experimental distribution and by using a Q-Q plot (see Figure \ref{qq}) we verify when there is a departure, statistical significant, from the straight line. Just to remind, a Q-Q plot shows the quantiles of one distribution as a function of the quantiles of another distribution. If the plot gives a straight line with a slope of 45$^\circ$, the two distributions are practically the same. Meaning that the theoretical distribution fits well the experimental distribution. Instead, a departure from the straight line means a not perfect concordance between the two distributions. 
In our case (Figure \ref{qq}, panel a) the WD is not suited for representing wind speed greater than 20 $m/s$, then, above this threshold we fit the empirical distribution with a GPD. From panel $b$ of the same figure, where the WD is used to model wind speed until threshold and GPD over threshold, one can notice the improvement in the fitting of the extreme values of the wind speed distribution with respect to the simple WD.

\begin{figure}
\centering
\includegraphics[height=8cm]{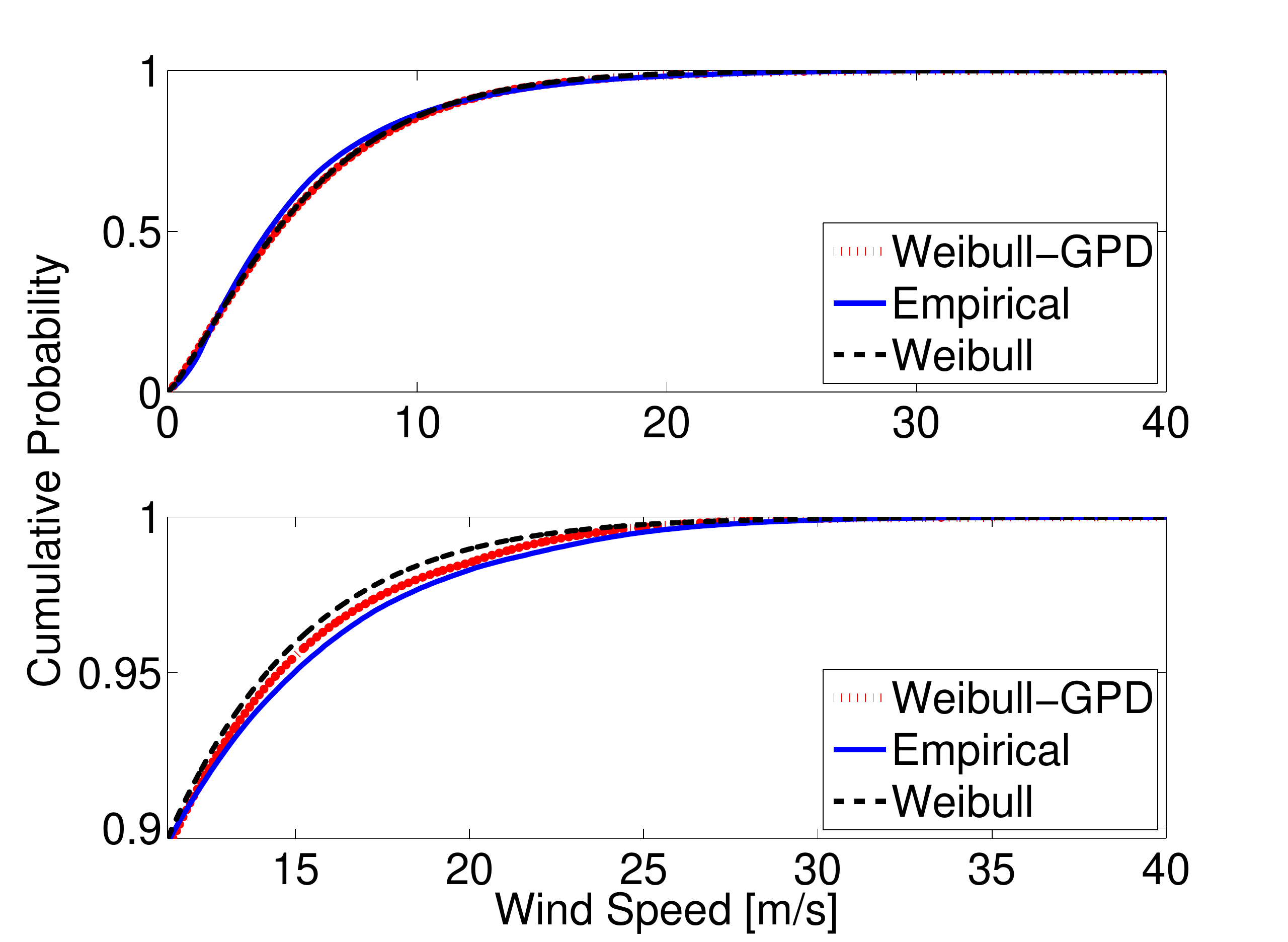}
\caption{Comparison of the cumulative distribution function of the Weibull distribution, Generalized Pareto distribution and of the real wind speed data. Upper panel: entire cumulative distribution function: lower panel: zoom on the tail of the distribution.}\label{cdf}
\end{figure}

Another comparison between the two distributions can be made by plotting their cumulative distribution function. In Figure \ref{cdf}, upper panel, we show the empirical cumulative distribution function of the real data (blue line), the Weibull (dashed black line) and the W-GP (red dotted line) cumulative distribution function. As it is possible to note in the lower panel of figure \ref{cdf}, the W-GP is closer to the empirical cumulative distribution function, with respect to the simple Weibull. From this figure we can highlight that if we model the wind speed with a WD we have an underestimation of the probabilities of events over a given threshold. As we will show better later, this implies that a model based on WD underestimates the energy produced by a wind farm due to the underestimation of high values of the wind speed.

\subsection{Copulas application and Monte Carlo simulation}

In the application tested here we consider a wind farm composed of 10 turbines. The correlations between the wind turbines that compose the wind farm are taken into account through Copulas. We consider two kind of Copulas, Gaussian and Student's t, the latter with different values of the degree of freedom (DOF). We consider constant values of the correlational matrix between the turbines and we study the variation on the production of energy by varying the correlation coefficients. We also tested the results with respect to the chosen Copula and to the DOF in the case of the Student's t Copula. The algorithm used for Monte Carlo simulation is described in section \ref{model}

\subsection{Wind turbine failures application}
\begin{figure}
\centering
\includegraphics[height=11cm]{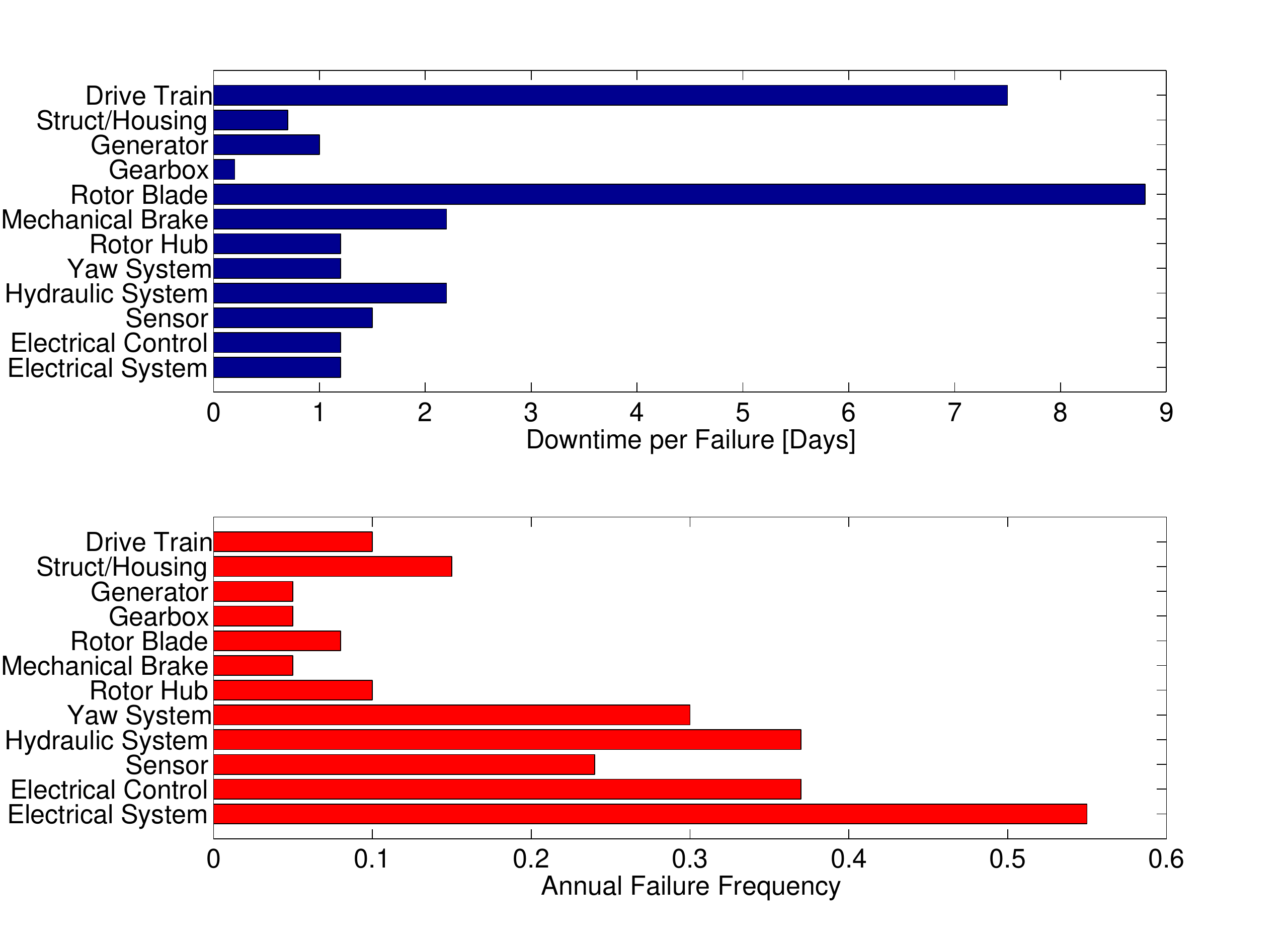}
\caption{Failure data for a wind farm composed of 69 Enercon E33 wind turbines. For each kind of failure are plotted the frequency per year and the time necessary to repair.}\label{poi}
\end{figure}

There are many reasons that can cause the absence of energy production in a wind farm derived from wind turbine failures. In figure \ref{poi} we summarize these kinds of failures and for each of them we show the annual frequency and the time necessary to repair the blades. Data are taken from \citet{tavn2010} and they are referred to a wind farm composed of 69 wind turbines, located in Ormont (Germany), of the same type of that chosen for our analysis (see section \ref{data}).

As mentioned in section \ref{model}, we assume that the number of wind turbine failures per year follows a Poisson distribution with parameter $\lambda=0.2$. This value represents the average of the number of wind turbine failures per year. Moreover we also assume that the repair time follows a Poisson distribution with parameter $\mu=2.4$. Also in this case $\mu$ is the average of the data of figure \ref{poi}, upper panel. In this way we can simulate the total time of no production by using the compound Poisson model.

\subsection{Energy conversion and sensitivity analysis}
\label{spie}

\begin{figure}
\centering
\includegraphics[height=8cm]{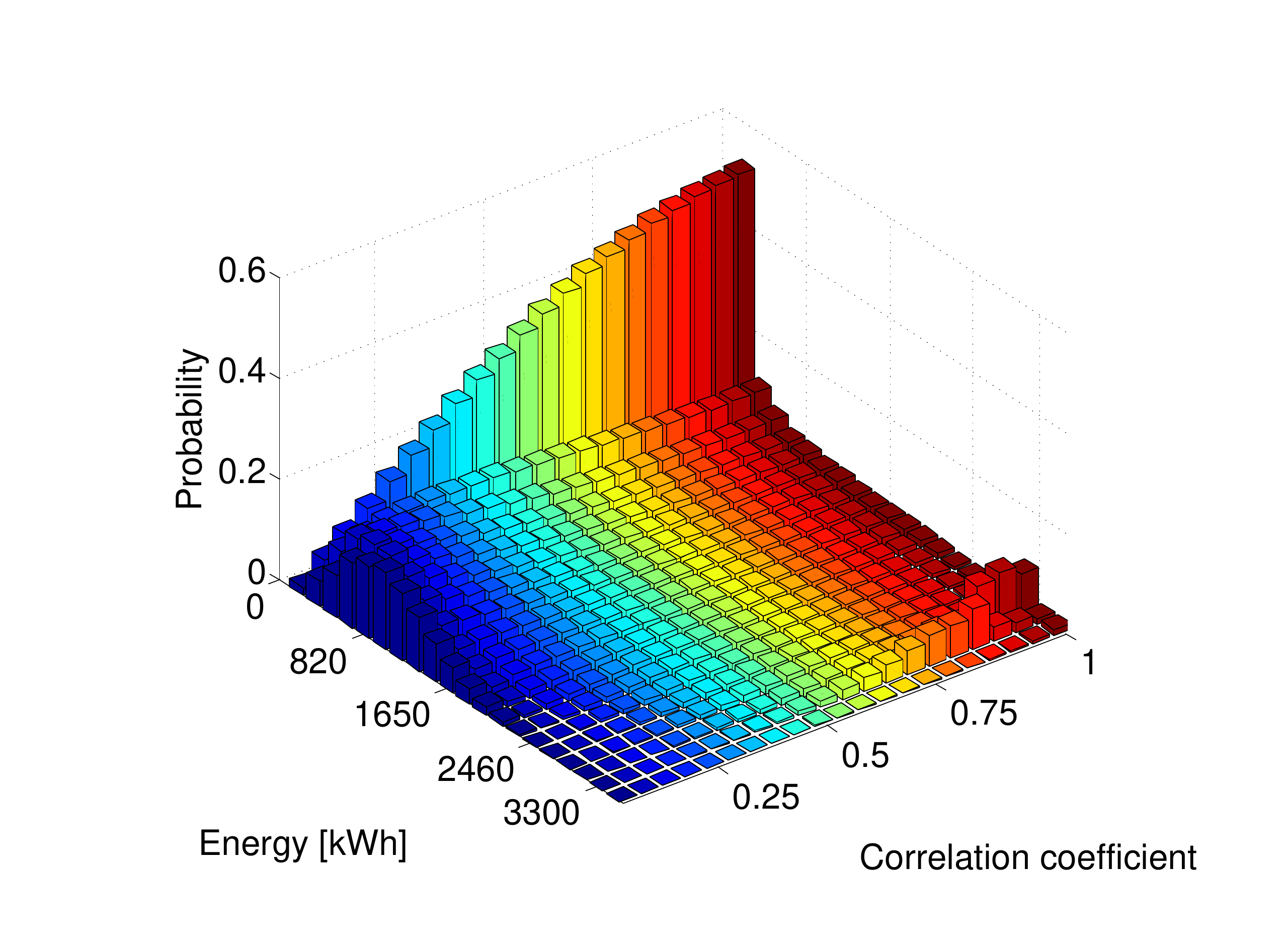}
\caption{Probability of energy production of the entire wind farm as a function of the correlation coefficient with a time interval of 6 minutes for a gaussian Copula.}\label{corr}
\end{figure}

At this point of the simulation we have 10 synthetic wind speed time series, one for each wind turbine, with a length equal to the number of years that we want simulate, then 10, correlated through Copula and with period of no production inside disposed randomly. Wind speed is converted into energy by using the power curve described in Section \ref{blades}.
In figure \ref{corr} we show the probability to produce a given energy every six minutes as a function of the correlation coefficients using a gaussian Copula. As it is possible to note, for low values of the correlation coefficient, the probability distribution becomes equal to a WD. Instead, for high values of the correlation coefficient, it assumes the common form of the energy production of a wind turbine, where the most likely values assumed by the system are in correspondence of the rated power and of absence of wind.

\begin{figure}
\centering
\begin{subfigure}[b]{0.45\textwidth}
\includegraphics[width=\textwidth]{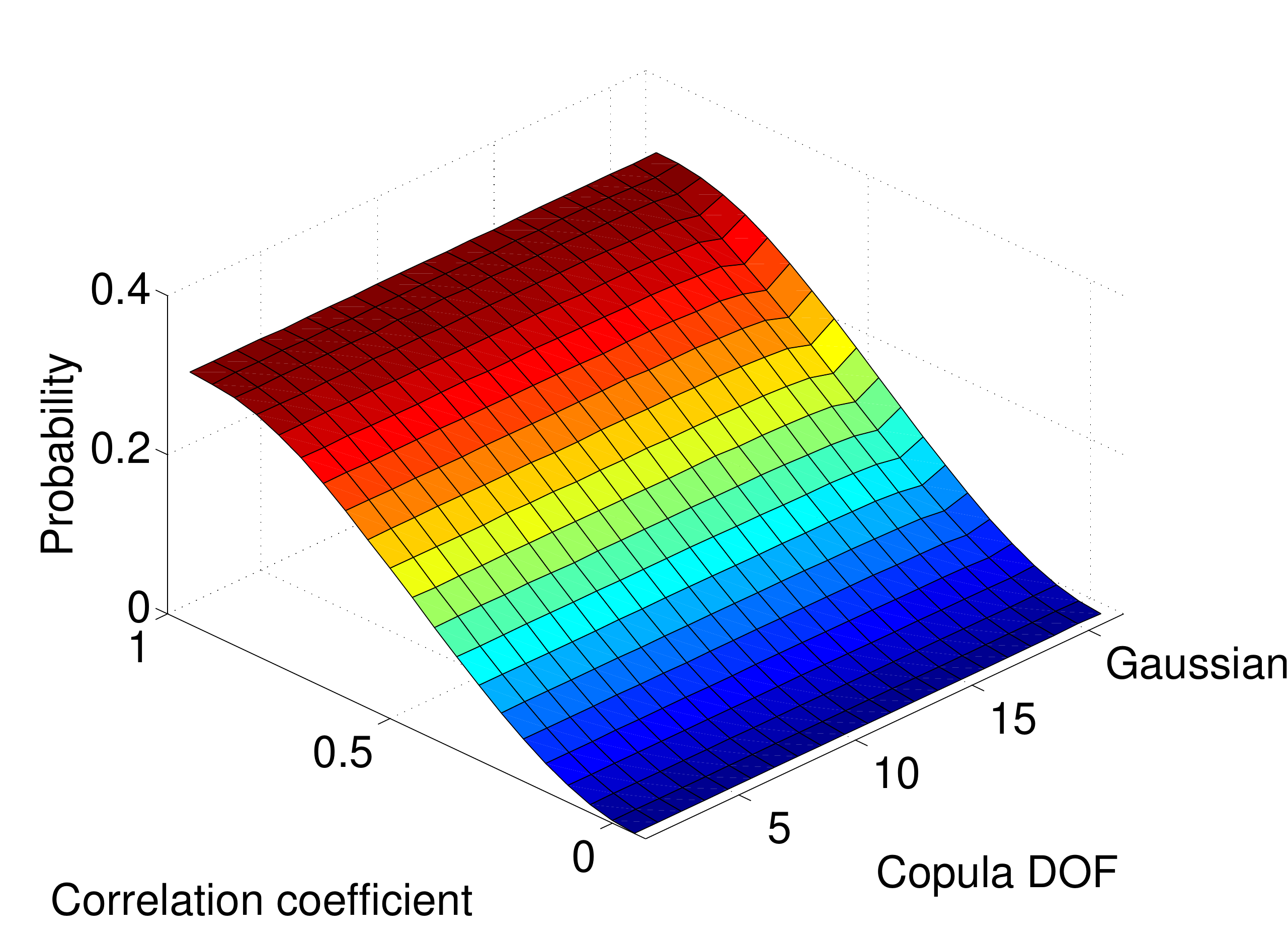}
\caption{}
\label{a1}
\end{subfigure}%
~ 
\begin{subfigure}[b]{0.45\textwidth}
\includegraphics[width=\textwidth]{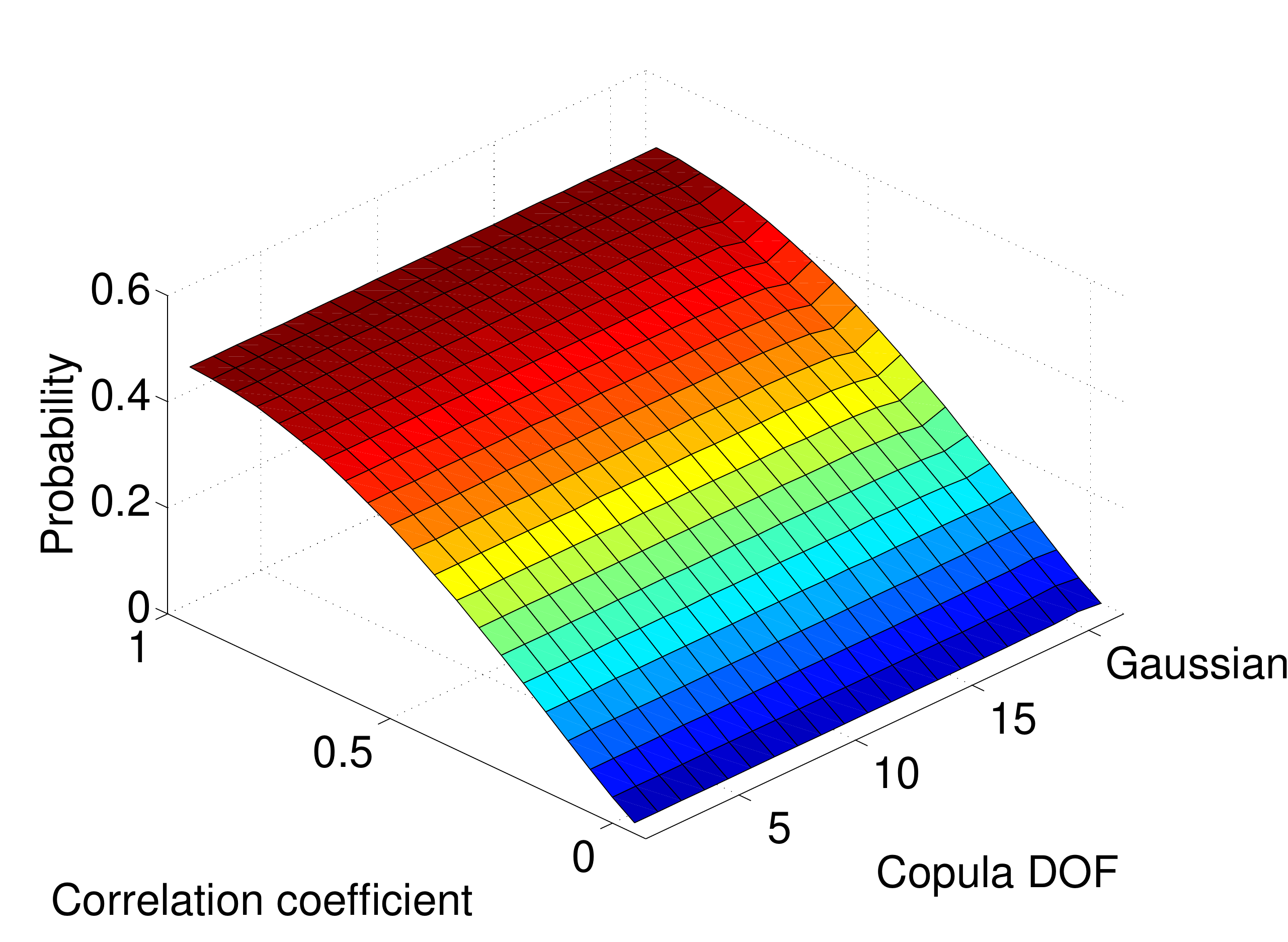}
\caption{}
\label{a2}
\end{subfigure}\\

\begin{subfigure}[b]{0.45\textwidth}
\includegraphics[width=\textwidth]{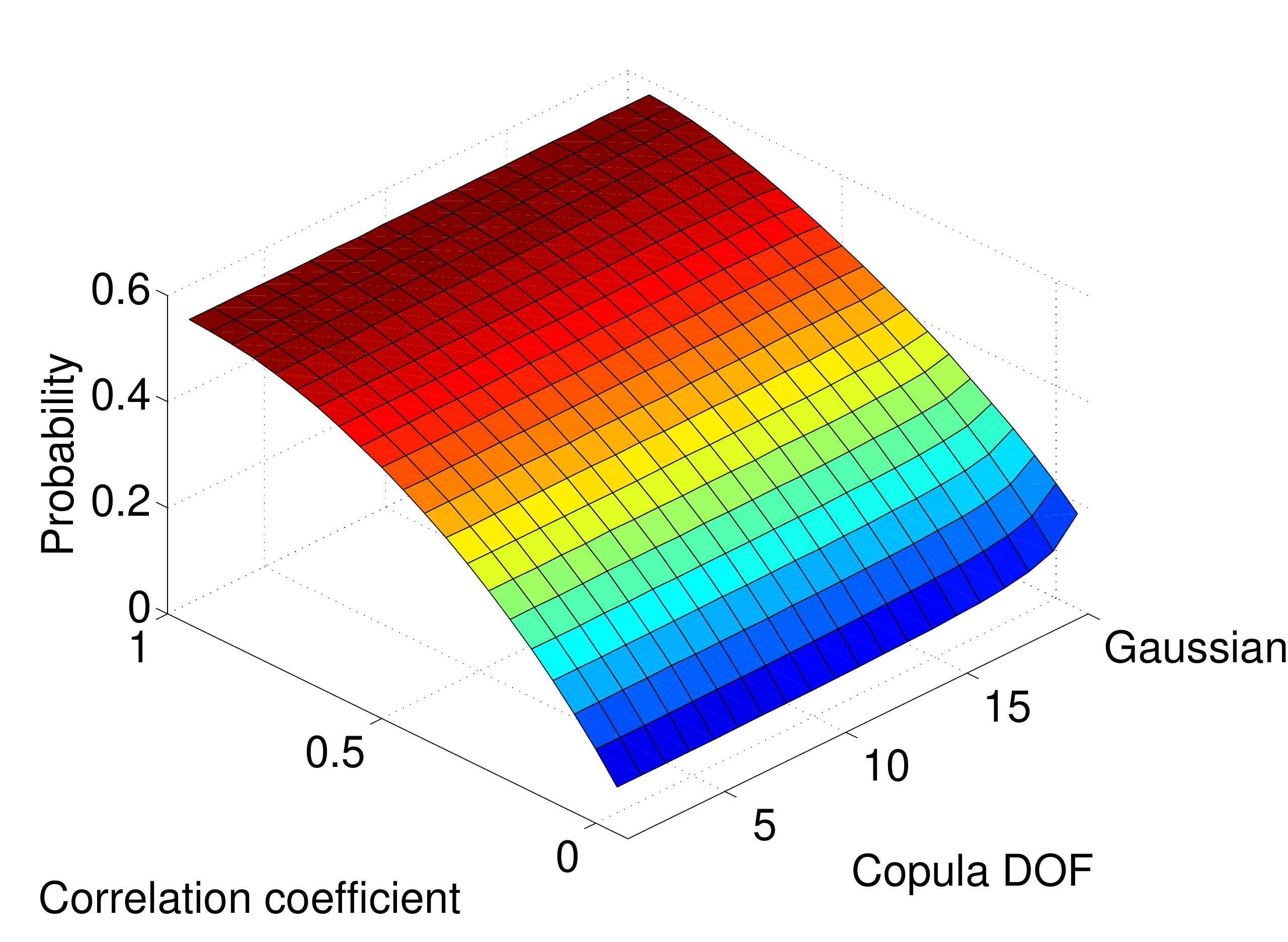}
\caption{}
\label{a3}
\end{subfigure}
~ 
\begin{subfigure}[b]{0.45\textwidth}
\includegraphics[width=\textwidth]{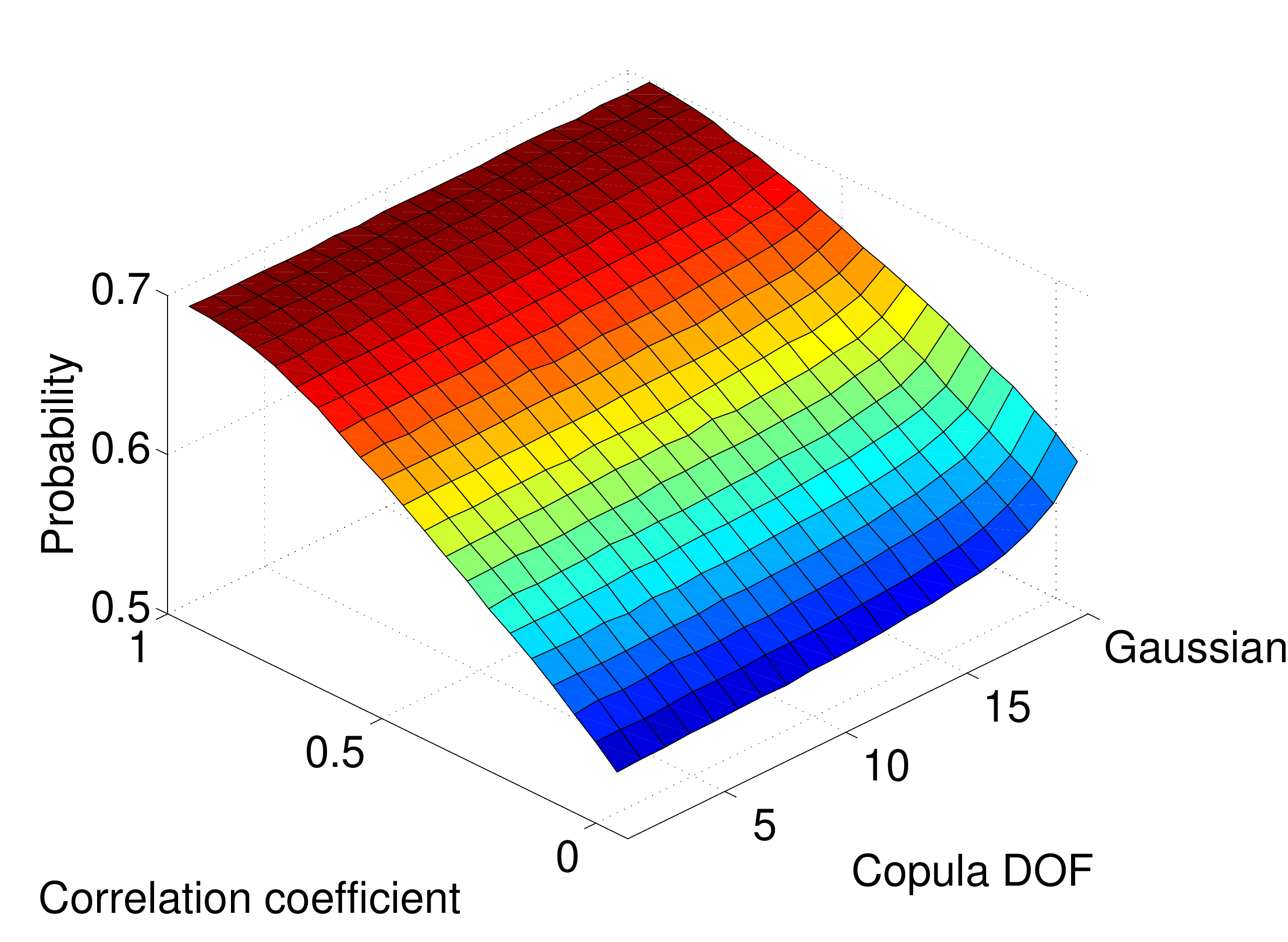}
\caption{}
\label{a4}
\end{subfigure}\\

\begin{subfigure}[b]{0.45\textwidth}
\includegraphics[width=\textwidth]{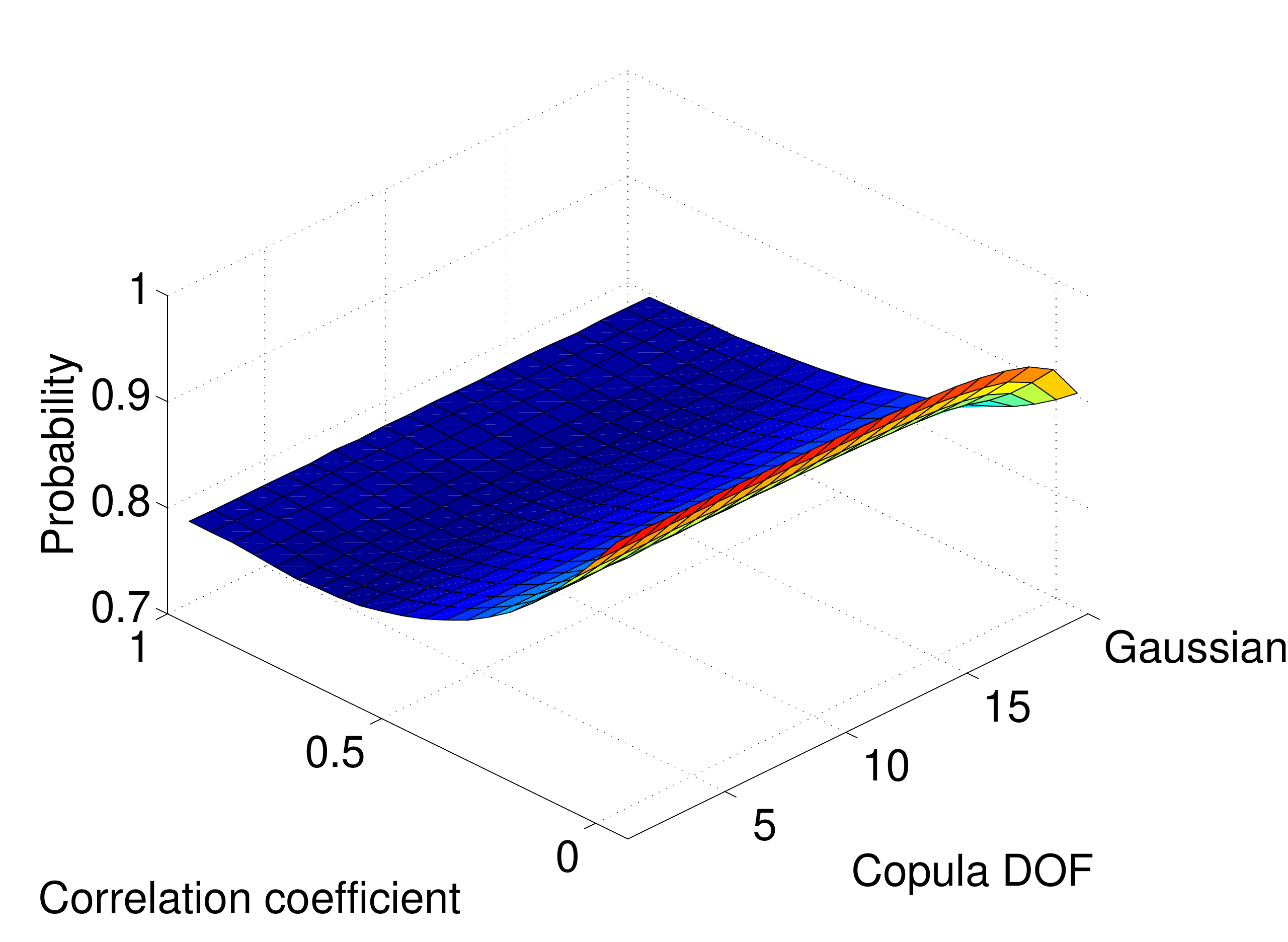}
\caption{}
\label{a5}
\end{subfigure}
~ 
\begin{subfigure}[b]{0.45\textwidth}
\includegraphics[width=\textwidth]{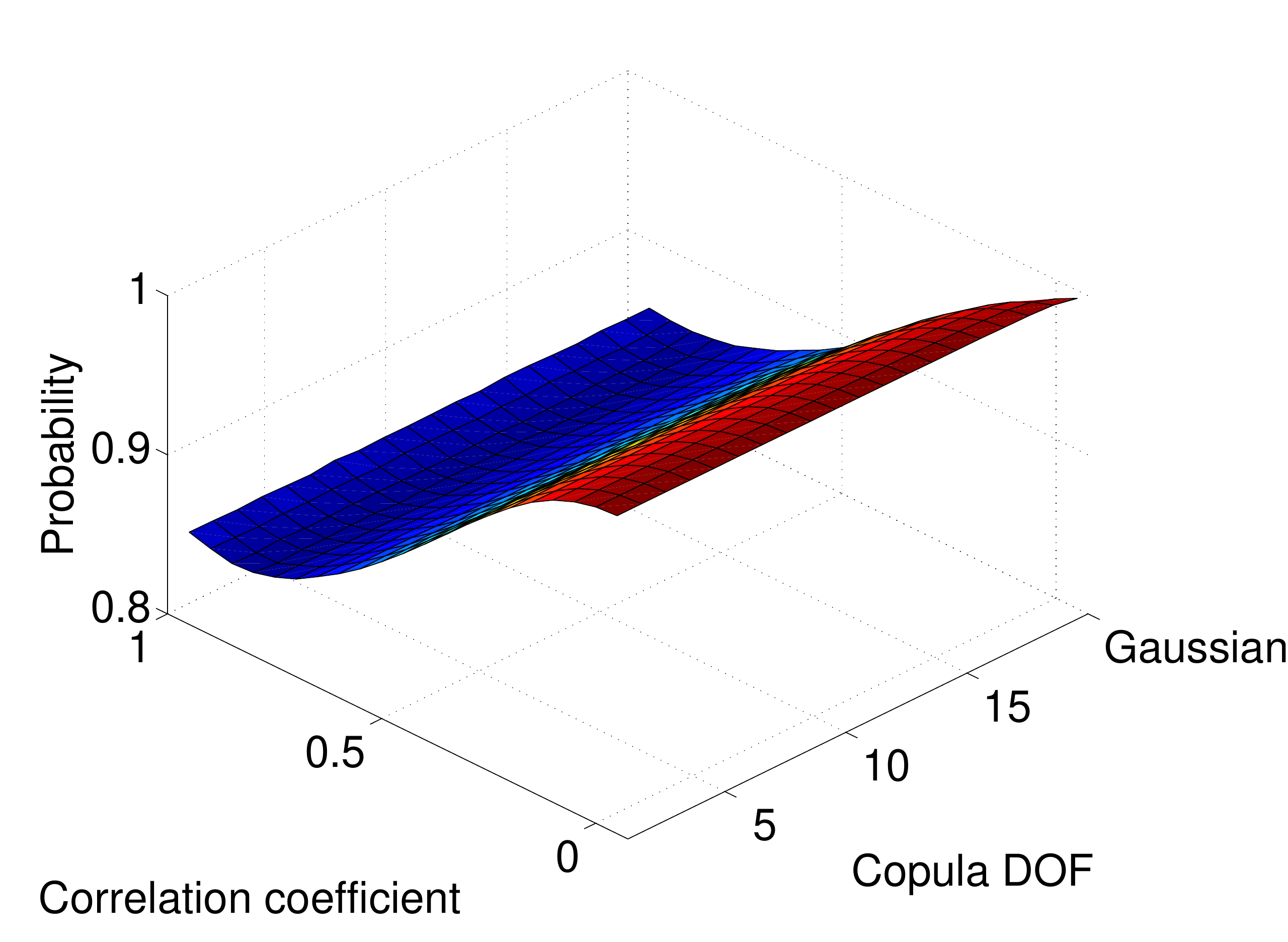}
\caption{}
\label{a6}
\end{subfigure}\\

\caption{Probability of producing under the (a) 1\%, (b) 5\%, (c) 10\%, (d) 25\%, (e) 50\% and (f) 75\% of the wind farm rated power as a function of the degree of freedom of the Copula and of the correlation coefficient.}\label{cor_cop}
\end{figure}

In Figure \ref{cor_cop} we show the probability of producing energy below a certain percentage of the rated power of the wind farm (the rated power of a wind farm is equal to the rated power of the wind turbine multiplied for the number of wind turbines that compose the wind farm, in this case it is equal to 3300 $kW$) as a function of the correlation coefficient and of the DOF of the Copula. Particularly, we show the probability of producing below 1\%, 5\%, 10\%, 25\%, 50\% and 100\% of the rated power respectively in panels, $a$, $b$, $c$, $d$, $e$ and $f$. 
As it is evident in the panels of figure \ref{cor_cop} there is a great variation of the probability with the variation of the correlation coefficient and a little dependence of the probability from the chosen Copula. Another important results is that the probability to produce less than a given threshold (at least when this threshold is below 50\% of the rated power) is much higher for correlated turbine than for uncorrelated ones. The results are opposite when the threshold is above 50\% of the rated power.

\begin{figure}
\centering
\includegraphics[height=9cm]{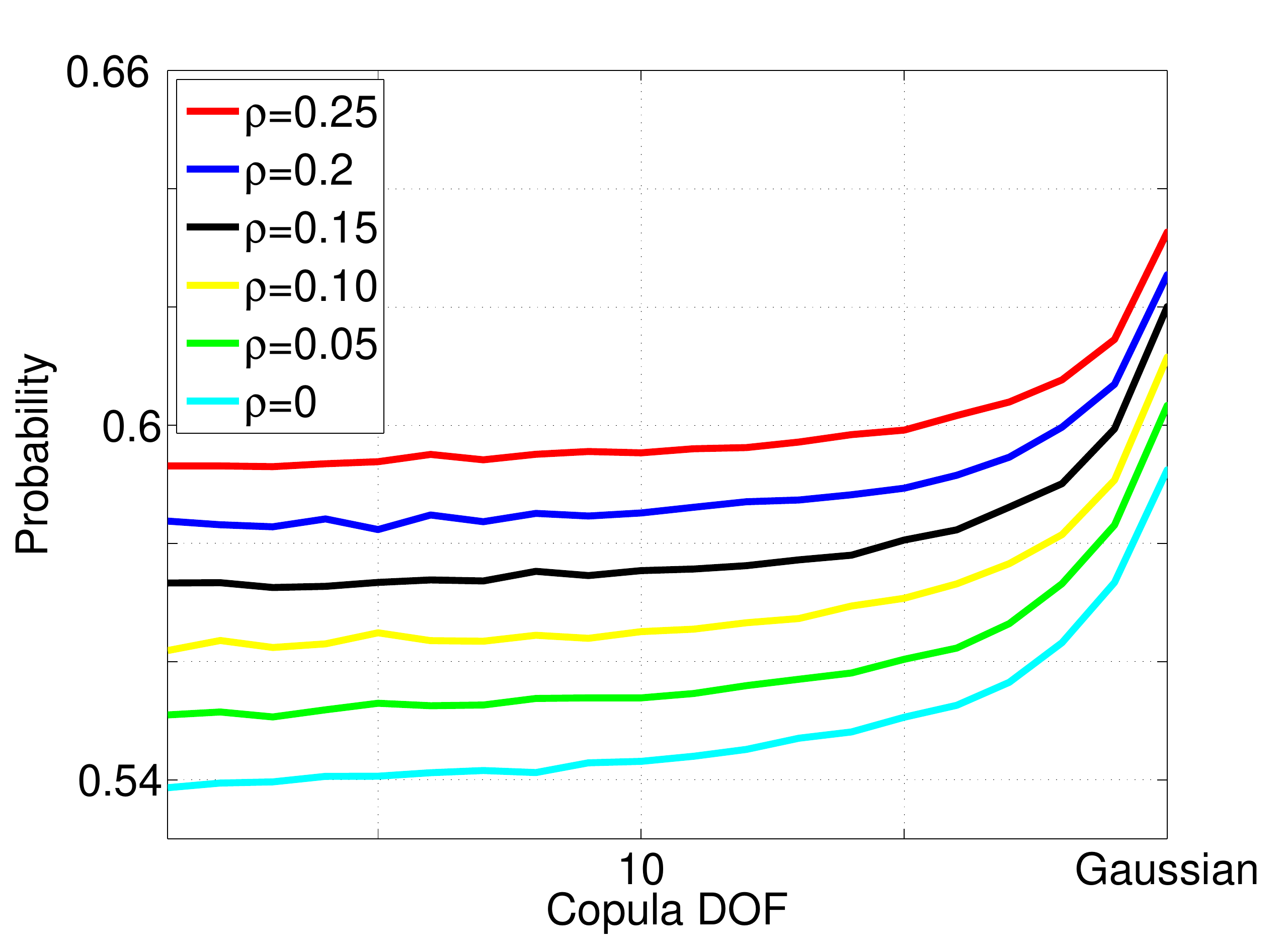}
\caption{Probability to produce less than 25\% of the wind farm rated power as a function of the degree of freedom of the Copula.}\label{vc}
\end{figure}

In order to highlight this phenomenon we fix the threshold at 25\% of the rated power and, for different values of the correlation coefficient, we plot the probability as a function of the DOF of the Copula. Results are shown in figure \ref{vc}, where it is possible to see that there is a little variation of the probability with the increase of the DOF of the Copula.

\subsection{Energy production: comparison with simple Weibull based model}
\label{lk}
\begin{figure}
\centering
\includegraphics[height=9cm]{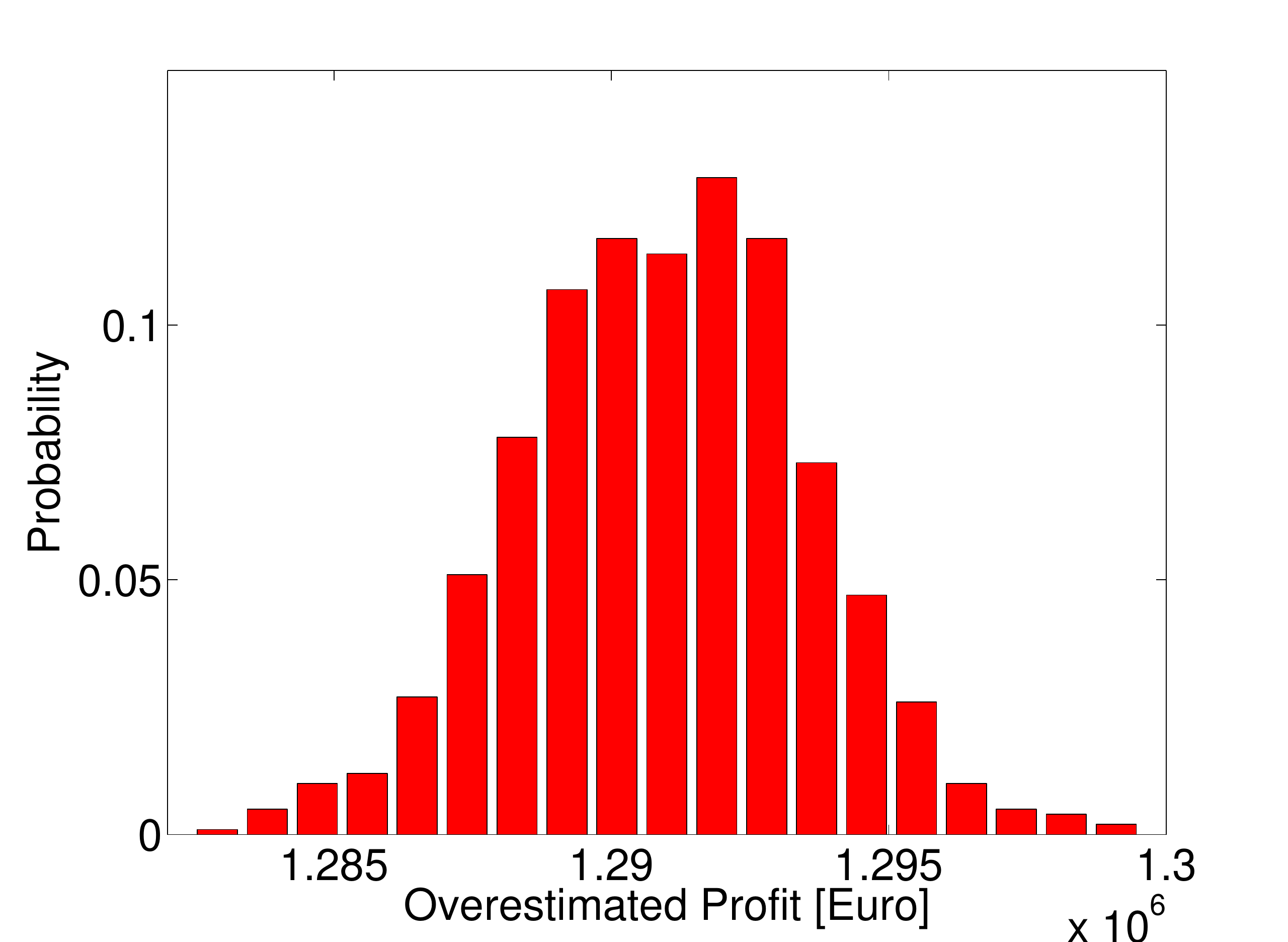}
\caption{Simulation of the difference between the energy produced (expressed in Euro) between the proposed procedure and a simple Weibull based model.}\label{wp}
\end{figure}

Lastly, we compared the estimated produced energy by using the proposed methods and the estimated produced energy by using a simple Weibul model for wind speed. As it was clear in the explanation of figure \ref{cdf} the WD underestimate the probability of wind speed exceeding a given threshold. This fact implies an underestimation in terms of energy produced by the entire wind farm. 
We simulate 10 distinct time series of wind speed (corresponding to 10 wind turbine composing the wind farm) with a length of 10 years. We transform the wind speed into energy produced and then in Euro by considering a price of 0.3 $Euro/kWh$. At last, we consider the difference between the energy estimated with the model presented in this paper and with the WD. This simulation was repeated 1000 times and the results are plotted in figure \ref{wp}. This Figure shows that, by using a simple WD model, we have a mean underestimation of production of about 1.29 millions of Euro in 10 years. This results is most interesting if we consider that with the new procedure we take into account also the lost of energy due to the wind turbines failures.

\section{Discussion and conclusion}
\label{res}

The goal of this paper is to propose a procedure to estimate accurately possible loss generated by the presence of period of no production caused by wind turbine failures and, on the other side, to fit well high values of the wind speed that can cause both no production or high production of energy, if the cut off wind speed of the wind turbine is exceeded or not. Nevertheless the use of the Copula appears as a good instruments to take into account the relationship between the wind turbines of the farm. The sensitivity analysis showed that the energy production of the wind farm is influenced by the DOF o the Copula but, above all, by the correlation coefficients.

Future work will focus on using the proposed procedure in order to evaluate new wind farm sites, to apply it to real case of investment into wind energy production and to extend the analysis when we assume a stochastic evolution of the wind speed process.

\bibliographystyle{elsarticle-harv} 
\bibliography{b}

\end{document}